\documentclass[prd,nofootinbib]{revtex4}
\usepackage{graphicx}
\usepackage{amsmath} 
\usepackage{amssymb}
\usepackage{bm}
\def\be{\begin{equation}}
\def\ee{\end{equation}}
\def\ba{\begin{eqnarray}}
\def\ea{\end{eqnarray}}

  \def\be{\begin{equation}}
\def\ee{\end{equation}}
 \def\bi{\begin{itemize}}
 \def\ei{\end{itemize}}
  \def\ben{\begin{enumerate}}
\def\een{\end{enumerate}}
  \def\bt{\begin{tabular}}
\def\et{\end{tabular}}
\def\bc{\begin{center}}
\def\ec{\end{center}}

\def\bea{\begin{eqnarray}}
\def\eea{\end{eqnarray}}
\def\l{\left}
\def\r{\right}
\def\f{\frac}
\def\hub{{\cal H}}

\def\bo{\bar{\omega}}

\def\mn{\mu\nu}

\begin{document}

\title{The pattern of growth in viable $f(R)$ cosmologies}
\author{Levon Pogosian$^{1}$}
\author{Alessandra Silvestri$^{2}$}

\affiliation{$^{1}$ Department of Physics, Simon Fraser University, Burnaby, BC, V5A 1S6, Canada,}
\affiliation{$^{2}$Department of Physics, Syracuse University, Syracuse, NY 13244, USA.}
\date{\today}

\begin{abstract}
We study the evolution of linear perturbations in metric $f(R)$ models of gravity and identify a potentially observable characteristic scale-dependent pattern in the behavior of cosmological structures. While at the background level viable $f(R)$ models must closely mimic $\Lambda$CDM, the differences in their prediction for the growth of large scale structures can be sufficiently large to be seen with future weak lensing surveys. While working in the Jordan frame, we perform an analytical study of the growth of structures in the Einstein frame, demonstrating the equivalence of the dynamics in the two frames. We also provide a physical interpretation of the results in terms of the dynamics of an effective dark energy fluid with a non-zero shear. We find that the growth of structure in $f(R)$ is enhanced, but that  there are no small scale instabilities associated with the additional attractive ``fifth force''. We then briefly consider some recently proposed observational tests of modified gravity and their utility for detecting the $f(R)$ pattern of structure growth.
\end{abstract}

\maketitle

%%%%%%%%%%%%%%%%%%%%%%%%%%%%%%%%%%%%%%%%%%%%%%%
\section{Introduction}
%%%%%%%%%%%%%%%%%%%%%%%%%%%%%%%%%%%%%%%%%%%%%%

Modern cosmology, with its many successes, is built on the concept of an expanding universe governed by the rules of Einstein's theory of General Relativity (GR). Assuming a homogeneous and isotropic ansatz for the metric, and given the material constituents, one can use GR to determine how the universe and its content evolve with time. It has gradually become apparent that, within this construction, ordinary particles and dark matter are not sufficient to describe what we observe. Current data strongly favor a universe that exponentially inflated at an early epoch and, after a period of decelerating expansion dominated by radiation and matter, recently began accelerating again. Within GR, a period of accelerated expansion can only be due to a dominating energy component with a negative equation of state, the most common example being a static, or slowly rolling, scalar field. Alternatively, one can try to avoid postulating new components by modifying GR. 

In 1979 Starobinsky showed that one can achieve a de Sitter phase in the early universe by replacing the Ricci scalar $R$ in the Einstein action with a function $R+{\rm const}\cdot R^2$ \cite{Starobinsky:1980te}. Starobinsky's model, which is now recognized as the first working model of Inflation, is also important as the first model for which the quantum metric fluctuations leading to a nearly scale-invariant initial density spectrum were discovered \cite{Mukhanov:1981}. More recently, $f(R)$ models have been revisited in the context of ongoing cosmic acceleration, starting with \cite{Capozziello:2003tk,Carroll:2003wy}
and followed by many others. 

The models proposed in \cite{Capozziello:2003tk,Carroll:2003wy} achieved late time acceleration by adding inverse powers of $R$ to the action, i.e. $R+{\rm const} \cdot R^{-n}$ with $n \ge 1$. It was later realized \cite{Dolgov:2003px,Amendola:2006kh,Nojiri:2006gh,Song:2006ej}, that these particular types of $f(R)$ models posses instabilities which, among other problems, prevent them from having a matter dominated epoch. Eventually, in \cite{Saw-Hu:2007,AmendolaPolarski,Agarwal:2007wn} conditions for cosmological viability of $f(R)$ theories were formulated and some explicit models satisfying them were constructed~\cite{Saw-Hu:2007,delaCruzDombriz:2006fj,Appleby,Starobinsky:2007hu,Nojiri:2007as}. In particular, it was explicitly demonstrated in \cite{Capozziello:2005ku,Nojiri:2006gh,Song:2006ej} that it is possible to ``design'' $f(R)$ functions with a stable matter era to match any desired expansion history of the universe.

It is well known that $f(R)$ theories can be recast as GR with an additional scalar field conformally coupled to all matter \cite{whitt84,maeda89,Magnano:1993bd,Chiba:2003ir}. Hence, they are a subset of a wider class of scalar-tensor theories with a fixed value of the coupling strength\footnote{It is perhaps a matter of personal choice whether or not to call $f(R)$ a ``modification of gravity'', or treat it as a class of models somehow separate from other scalar-tensor theories. At least from the model-building point of view, they are as a class of models that operationally do not require postulating a dark energy component.}. Since this coupling is of order one in Planck units, it would seem that all $f(R)$ models would be automatically ruled out for violating the stringent equivalence principle tests (EPT) in the solar system and on Earth \cite{CMWill}. However, just like in the case of Chameleon models \cite{Khoury:2003aq,Khoury:2003rn}, the $f(R)$ scalar degree of freedom can be exponentially suppressed in regions of high matter density. Hence, while the EPT impose strong constraints on $f(R)$, they do not automatically rule them out \cite{Cembranos:2005fi,Saw-Hu:2007}. In Section~\ref{viable} we show that the EPT constraints only allow $f(R)$ models that are practically indistinguishable from the cosmological constant at the background level. This does not necessarily mean that there cannot be differences in the dynamics of perturbations. In fact, as already shown in \cite{Saw-Hu:2007}, and as we also demonstrate in this Paper, $f(R)$ can lead to interesting signatures that can be seen with future weak lensing surveys.

Linear perturbations in $f(R)$ models have already been studied in \cite{Song:2006ej,Bean:2006up,Li:2007xn,Tsujikawa:2007gd} using differing  analytical and numerical techniques. The results were obtained in either Jordan or Einstein frames, and for particular choices of f(R) models. In this work, we identify the general features of perturbation dynamics common to all viable $f(R)$ models and connect the results obtained in different frames. In addition, we show that one can interpret the $f(R)$ perturbations in terms of the dynamics of an effective dark energy fluid with a non-zero shear. We then demonstrate the signature of $f(R)$ as would be seen via some recently proposed observational tests aimed at detecting the cosmic shear \cite{Caldwell,Bean-Dod}.

We start with a review of $f(R)$ theories and the conditions for their viability in Section \ref{fofrdescription}. In section \ref{perttheory} we solve the full set of linear perturbation equations in the Jordan frame in the Newtonian gauge and present the numerical solutions, as well as an analytical study of the sub-horizon perturbations in both Jordan and Einstein frames, and in terms of an effective dark energy fluid with a non-zero shear. We conclude with Section \ref{conclusions}, where we discuss the prospects of observationally detecting the $f(R)$ pattern of structure growth.

%%%%%%%%%%%%%%%%%%%%%%%%%%%%%%%%%%%%%%%%%%%%%%%
\section{$f(R)$ gravity models}
\label{fofrdescription}
%%%%%%%%%%%%%%%%%%%%%%%%%%%%%%%%%%%%%%%%%%%%%%%

We focus on gravity theories described by the action
\be
S=\frac{1}{2\kappa^{2}}\int d^4 x\sqrt{-g}\, \left[R+f(R)\right] + \int d^4 x\sqrt{-g}\, {\cal L}_{\rm m}[\chi_i,g_{\mu\nu}] \ ,
\label{jordanaction}
\ee
where $\kappa^{2}= 8\pi G$ and $f(R)$ is a general function of the Ricci scalar, $R$. The matter Lagrangian, ${\cal L}_{\rm m}$, is minimally coupled and, therefore, the matter fields, $\chi_i$, fall along geodesics of the metric $g_{\mu\nu}$. The field equations obtained from varying the action~(\ref{jordanaction}) with respect to $g_{\mu\nu}$ are
\be \label{jordaneom}
\left(1+f_R \right)R_{\mu\nu} - \frac{1}{2} g_{\mu\nu} \left(R+f\right) + \left(g_{\mu\nu}\Box 
-\nabla_\mu\nabla_\nu\right) f_R =\kappa^{2}T_{\mu\nu} \ ,
\ee
where we have defined $f_R\equiv \partial f/\partial R$. We take the energy-momentum tensor to be that of a perfect fluid,
\begin{equation}
T_{\mu\nu} = (\rho+ P)U_{\mu} U_{\nu} + P g_{\mu\nu}\ ,
\label{perfectfluid}
\end{equation} 
where $U^{\mu}$ is the fluid rest-frame four-velocity, $\rho$ is the energy density and $P$ is the pressure. $P$ is related to $\rho$ via $P=w\rho$, where $w$ is the equation of state parameter (for pressureless matter $w=0$ and for radiation $w=1/3$). When considering the background cosmological evolution, we take the metric to be of the flat Friedmann-Robertson-Walker form, $ds^2 = a^{2}(\tau)(-d\tau^{2} + d{\bf x}^2)$, where $\tau$ is the conformal time and $a(\tau)$ is the scale factor.
The metric $g_{\mu\nu}$ is minimally coupled to matter, hence the stress tensor and its conservation law will be the same as in the standard GR. In particular, the continuity equation is the usual 
\ba
\dot{\rho}+3\hub(\rho+P) &=& 0 \ ,
\label{jordancontinuity}
\ea
%, in terms of which the curvature scalar satisfies $R =6 \ddot{a}/a^{3}$. 
where an overdot denotes differentiation with respect to conformal time and $\hub \equiv \dot{a} / a$.
%is the corresponding Hubble expansion rate.  
The $f(R)$ term in the gravitational action leads to extra terms in the Einstein equations, which now become fourth order differential equations (in contrast to the second order of the equations in standard GR). In particular, for our cosmological ansatz, the Friedmann equation becomes
\ba
(1+f_{R})\hub^2+\frac{a^{2}}{6}f-\f{\ddot{a}}{a}f_{R} +\hub \dot{f}_{R} &=&  \f{\kappa^{2}}{3}a^{2}\rho
\label{jordanfriedmann}
\ea
and the acceleration equation is
\be\label{acceleration}
\f{\ddot{a}}{a}-(1+f_R)\hub^2+a^2\f{f}{6}+\f{1}{2}\ddot{f_R}=-\f{\kappa^2}{6}a^2(\rho+3P)
\ee
One can interpret the extra terms in the Einstein equations (\ref{jordaneom}) as a contribution of an {\it effective fluid} with energy-momentum tensor
\be\label{G_eff}
\kappa^2{T}^{\rm{eff}}_{\mn} \equiv f_RR_{\mn}-\f{1}{2}fg_{\mn}+(g_{\mn}\Box-\nabla_{\mu}\nabla_{\nu})f_R \ .
\ee
The effect of this fluid on the cosmological background can be described in terms of an effective density
\be\label{enden_eff}
\rho_{\rm{eff}}=\f{1}{\kappa^2}\l[\f{1}{2}(f_R R-f)-3\f{\hub^2}{a^2}f_R-3\f{\hub}{a^2}\dot{f}_R\r]\ ,
\ee
and the equation of state 
\be\label{eos_eff}
w_{\rm{eff}}=-\f{1}{3}-\f{2}{3}\f{\l[\hub^2f_R-\f{1}{6}a^2f-\f{1}{2}\hub \dot{f}_R-\f{1}{2}\ddot{f}_R\r]}{\l[-\hub^2f_R-\f{1}{6}a^2f-\hub \dot{f}_R+\f{1}{6}a^2f_RR\r]} \ .
\ee
This effective fluid approach can facilitate the comparison with Dark Energy (DE) models, as well as the interpretation of the effects of modifications to GR. It is, however, important to keep in mind that the extra terms associated with modifications to GR have a geometrical origin.

By definition, $f(R)$ models with the same matter content and the same $w_{\rm{eff}}$ will have identical background evolution histories. In fact, for any given function $w_{\rm{eff}}(a)$ one can solve the differential equation (\ref{eos_eff}) and find a family of different $f(R)$ with the same expansion history but corresponding to different choices of boundary conditions. In particular, for $w_{\rm{eff}}=-1$, eq.~(\ref{eos_eff}) leads to
\be
\label{lambda_eos}
\ddot{f}_R-2\hub\dot{f}_R+2\l(\dot{\hub}-\hub^2\r)f_R=0 \ .
\ee
One can then set $\hub^2=(\kappa^2 a^2/3)[\rho + \rho_{\rm{eff}}]$
and integrate eq.~(\ref{lambda_eos}) to find all $f(R)$ that have a $\Lambda$CDM background expansion history. Analogous equations can be written for any function $w_{\rm{eff}}(a)$.

The role of DE in eqs.~(\ref{jordanfriedmann}) and (\ref{acceleration}) is effectively played by the additional terms due to $f(R)\ne 0$. There exists, however, a complementary, and sometimes conceptually simpler, way in which to approach $f(R)$ theories. It is possible to conformally transform the metric and bring the gravitational part of the action to the usual Einstein-Hilbert form of standard GR.  The price one pays for this simplification is a non-minimal coupling between matter fields and the transformed metric \cite{Cotsakis:1988,Amendola:1999er,Bean:2000zm}, as well as the appearance of a new scalar degree of freedom playing the role of DE and evolving under a potential determined by the original form of the $f(R)$ in (\ref{jordanaction}). Assuming the transformation between the two frames is always well-defined, the results obtained using either description should be the same. The frame in which matter particles fall along the geodesics of the metric, while the form of the gravitational part of the action may be modified, is referred to as the \emph{Jordan frame}. The frame in which the gravitation part of the action is the same as in standard GR, while the matter may be non-minimally coupled to gravity, is called the \emph{Einstein frame}. In the present paper we primarily work in the Jordan frame. When discussing the perturbations, we will make the connection between the Jordan and Einstein frame descriptions. The description of the Einstein frame version of $f(R)$ theories and the mapping between the two frames can be found in Appendix \ref{conformal}.

The emergence of an additional scalar degree of freedom can be seen directly in the Jordan frame. The role of the scalar field is played by $f_R$, dubbed {\it scalaron} in \cite{Starobinsky:1980te}. Indeed, the trace of eq~(\ref{jordaneom}) can be written as
\be
\Box{f}_R={1\over 3}\left(R+2f-Rf_R \right)-{\kappa^2 \over 3}(\rho-3P) 
\equiv {\partial V_{\rm{eff}} \over \partial f_R} \ ,
\ee
which is a second order equation for a field $f_R$ with a canonical kinetic term and an effective  potential $V_{\rm{eff}}(f_R)$. By design, the $f(R)$ theories we consider must have $|f\ll R|$ and $|f_R|\ll 1$ at high curvatures to be consistent with our knowledge of the high redshift universe. In this limit, the extremum of the effective potential lies at the GR value $R=\kappa^2(\rho-3P)$. Whether this extremum is a minimum or a maximum is determined by the second derivative of $V_{\rm{eff}}$ at the extremum, which is also the squared mass of the scalaron:
\be
m^2_{f_R} \equiv {\partial^2 V_{\rm{eff}} \over \partial f_R^2} = 
%{1\over f_{RR}}{\partial \over \partial R}\left[{\partial V_{\rm{eff}} \over \partial f_R}\right]=
{1\over 3}\left[{1+f_R \over f_{RR}}-R \right] \ .
\ee
At high curvatures, when $|Rf_{RR}|\ll 1$ and $f_R\rightarrow 0$, 
\be
m^2_{f_R} \approx {1+f_R \over 3f_{RR}} \approx {1 \over 3f_{RR}}\ .
\ee
It then follows that in order for the scalaron not to be tachyonic one must require $f_{RR}>0$. Classically, $f_{RR}>0$ is required to keep the evolution in the high curvature regime stable against small perturbations \cite{Dolgov:2003px,Sawicki:2007tf}. 
The scalaron mediates an attractive ``fifth force'', which has a range determined by the Compton wavelength
\be\label{lambda_C}
\lambda_C \equiv {2\pi \over m_{f_R}} \ .
\ee
While $\lambda_C$ is large at current cosmological densities, terrestrial, solar and galactic tests are not necessarily violated because the scalaron acquires a larger mass in regions of high matter density. This is essentially the Chameleon mechanism of \cite{Khoury:2003aq,Khoury:2003rn}.

\subsection{Designer $f(R)$}

As mentioned earlier, the $4$th order nature of $f(R)$ theories provides enough freedom to reproduce any cosmological background history by an appropriate choice of the $f(R)$ function \cite{Multamaki:2005zs,Capozziello:2006dj,Song:2006ej}.  Instead of using eq.(\ref{eos_eff}), we integrate the modified Friedmann equation (\ref{jordanfriedmann}) directly to find $f(R)$ models with specified expansion histories. In this subsection we describe this ``designer'' procedure, which generalizes that of  \cite{Song:2006ej} to include a radiation component and to allow for a time-dependent $w_{\rm{eff}}(a)$. 

As in \cite{Song:2006ej}, let us introduce dimensionless variables
\be
\label{dimensionless}
y \equiv {f(R) \over H_0^2} \ , \ \ E \equiv {H^2 \over H^2_0} \ ,
\ee
where $H\equiv a^{-1}da/dt$, $t$ is the physical time, and $H_0$ is the Hubble parameter today. We fix the desired expansion history to that of a flat universe containing matter, radiation and dark energy with a given equation of state. Namely, we set
\be
\label{E(x)}
E=\Omega_m a^{-3} + \Omega_r a^{-4} + \rho_{\rm{eff}}/\rho_c^0\equiv E_m+E_r+E_{\rm{eff}} \\
\ee
with
\be
\label{rhoeff}
E_{\rm{eff}}= (1-\Omega_m-\Omega_r)\exp\left[-3\ln a+3\int_a^1 w_{\rm{eff}}(a) d\ln a \right] \ ,
\ee
where $\rho_c^0 \equiv 3H^2_0/\kappa^2$ is the critical density today, $\Omega_{m,r}$ are present day fractions of matter and radiation densities and we have defined $E_i\equiv\rho_i/\rho_{cr}^0$ (that will be used later in the perturbation equations). The other background functions that appear in the modified Friedmann equation (\ref{jordanfriedmann}) can be expressed in terms of $E$ and its derivatives. For example,
\be
{R \over H_0^2} = 3(4E+E') \ .
\ee
where the prime denotes differentiation with respect to $\ln a$. Substituting (\ref{E(x)}) into eq.(\ref{jordanfriedmann}) yields a second order equation for $y$:
\be
y''-\left( 1+{E'\over 2E} + {R''\over R'}\right)y'+{R' \over 6H_0^2E}y =- {R' \over H_0^2E} {E_{\rm{eff}}} \ .
\label{y2prime}
\ee
In order to set the initial conditions, let us consider the general and the particular solutions of (\ref{y2prime}) at early times, when the effects of the effective dark energy on the expansion are negligible. At a certain early value $\ln a_i$, the homogeneous part of (\ref{y2prime}) is satisfied by a power law ansatz $y \propto a_i^p$. Substituting this ansatz in and solving the quadratic equation for $p$ yields
\be
p_\pm={1\over 2} \left(-b\pm \sqrt{b^2-4c}\right) \ ,
\ee
where
\be
b={7+8 r_i \over 2(1+r_i)} \ , \ \ c=-{3\over 2 (1+r_i)} \ , \ \ r_i={a_{eq} \over a_i} \ ,
\ee
and $a_{eq}$ is the scale factor at the radiation-matter equality.
The decaying mode solution corresponding to $p_-$ leads to a large $f(R)$ at early times which makes it unacceptable, and we set its amplitude to zero. The particular solution at $\ln a_i$ can be found by substituting
\be
y_p = A_p E_{\rm{eff}}(a_i) 
\ee
into (\ref{y2prime}), where $E_{\rm{eff}}$ is given by (\ref{rhoeff}). One then finds
\be
A_p={-6c \over -3w_{\rm{eff}}'+9w_{\rm{eff}}^2+(18-3b)w_{\rm{eff}}+9-3b+c} \ .
\ee
Put together, the initial conditions at $\ln a_i$ are
\ba
y_i&=&A e^{p_+\ln a_i} + y_p \\
\nonumber
y_i'&=&p_+A e^{p_+\ln a_i} - 3[1+w_{\rm{eff}}(a_i)] y_p \ ,
\label{eq:initial}
\ea
and $A$ is the remaining arbitrary constant that can be used to parametrize different $f(R)$ models with the same expansion history.

%Solving eq~(\ref{y2prime}) is  equivalent to (\ref{lambda_eos}) for $w_{\rm{eff}}=-1$.
%As we will argue in the next Subsection, only models with $w_{\rm{eff}}\approx -1$ can satisfy all of %the existing constraints. In addition, we illustrate in Section~\ref{perttheory} that linear %perturbations in models with $w_{\rm{eff}}=-1$ and those with $w_{\rm{eff}}$ very close to $-1$ can be made %essentially the same by an appropriate choice of $A$ in eq~(\ref{eq:initial}). Hence, for the %purpose of studying cosmological perturbations in viable $f(R)$ theories, it will be sufficient to %only consider models with $w_{\rm{eff}}=-1$

\subsection{Viable models}
\label{viable}

While it is possible to find $f(R)$ models that reproduce any given expansion history, they must satisfy additional conditions to be consistent with existing experimental and observational data.  In this subsection we summarize these requirements and examine their implications for the acceptable range of values of $w_{\rm{eff}}$.

The most common argument against $f(R)$ theories is based on their identification with scalar-tensor gravities \cite{whitt84,maeda89,Magnano:1993bd,Chiba:2003ir}. In particular, it has been shown that in the metric formalism considered here, $f(R)$ corresponds to a Brans-Dicke theory with $\omega_{BD}=0$ \cite{Chiba:2003ir}. Since low values of $\omega_{BD}$ have long been ruled out by solar system tests, it would appear that any $f(R)$ theory is automatically ruled out. However, it is also known \cite{Khoury:2003aq,Khoury:2003rn,Saw-Hu:2007} that in regions of high concentration of matter, the non-minimally coupled scalar degree of freedom acquires additional mass and, therefore, can be strongly suppressed. In such situations, simple arguments based on the formal equivalence with Brans-Dicke are insufficient to make conclusive statements about viability of $f(R)$ theories. Instead, one can formulate a set of requirements for these theories to satisfy in order to pass all of the tests. While these requirements are quite strict and generally discourage large deviations of $f(R)$ from the trivial case of a cosmological constant ($f=-2\Lambda$), there can be room for non-trivial observable differences. Such differences are interesting to understand in some detail and work has already been done on this in \cite{Song:2006ej,Tsujikawa:2007gd,Song:2007da}

Below we list and discuss the conditions that $f(R)$ models of cosmic acceleration must satisfy in order to be viable.
\begin{enumerate}
\item \label{cond:1} $f_{RR}>0$ for $R\gg f_{RR}$. Classically, this follows from requiring the existence of a stable high-curvature regime, such as the matter dominated universe \cite{Dolgov:2003px}. Quantum mechanically, $m^{-2}_{f_R} \approx f_{RR}>0$ ensures that the scalaron is non-tachyonic. 

\item \label{cond:2} $1+f_R>0$ at all finite $R$. The most direct interpretation of this condition is that the effective Newton constant, $G_{\rm{eff}}=G/(1+f_R)$, is not allowed to change sign. Among classical implications of $1+f_R<0$ is the universe quickly becoming inhomogeneous and anisotropic \cite{Nariai:1973eg,Gurovich:1979xg}. The quantum mechanical significance of this condition is in preventing the graviton from turning into a ghost \cite{Nunez:2004ji}. 

\item \label{cond:3} $f_R<0$. Given tight constraints from Big Bang nucleosynthesis (BBN) and  the Cosmic Microwave Background (CMB), we want regular GR to be recovered at early times, i.e. $f(R)/R$ and $f_R \rightarrow 0$ as $R\rightarrow \infty$. Together with $f_{RR}>0$, this implies that $f_R$ must be a negative, monotonically increasing function of $R$ that asymptots to $0$ from below.

\item \label{cond:chameleon} $f_R$ must be small at recent epochs. This is not required if the only aim is to build a model of cosmic acceleration, without trying to satisfy the solar and galactic scale constraints. This condition ensures a small difference between the value of the scalaron field in the high density galactic center and the value in the outskirts of the galaxy \cite{Saw-Hu:2007}. As shown in \cite{Khoury:2003rn}, the difference between these values is effectively the potential difference that sources the attractive ``fifth'' force acting on objects in the vicinity of the galaxy. Analysis of \cite{Saw-Hu:2007} suggests that the value of $|f_R|$ today should not exceed $10^{-6}$. However, as stressed in \cite{Saw-Hu:2007}, this bound assumes that in $f(R)$ models galaxy formation proceeds similarly to that in GR. Hence, while it is certain that $|f_R|$ must be small, any specific bound on its value today will be unreliable until galaxy formation in $f(R)$ is studied in N-body numerical simulations.

\end{enumerate}

These restrictions reduce the space of allowed values of $w_{\rm{eff}}$. Namely, there are functions $w_{\rm{eff}}(a)$ for which none of the solutions of eq.(\ref{y2prime}) satisfy all of the above conditions. To understand this better, let us examine a typical solution of eq.(\ref{y2prime}) for a given expansion history. It can be written as a sum of the solution of the homogeneous part of the equation plus the particular solution. During the radiation and matter domination the homogeneous part of eq.(\ref{y2prime}) is only weakly dependent on $w_{\rm{eff}}$. In that regime, the main impact of $w_{\rm{eff}}(a)$ is through the driving term on the RHS of eq.(\ref{y2prime}) which sets the time dependence of the particular solution. To gain intuition into which $w_{\rm{eff}}$ are viable, one could first ``turn off'' the homogeneous part by setting $A=0$ in (\ref{eq:initial}) and see if the above Conditions are satisfied. Then, if any of the Conditions are violated, one could try to remedy the situation by a suitable choice of $A$. 

\begin{figure}[h]
\includegraphics[width=100mm]{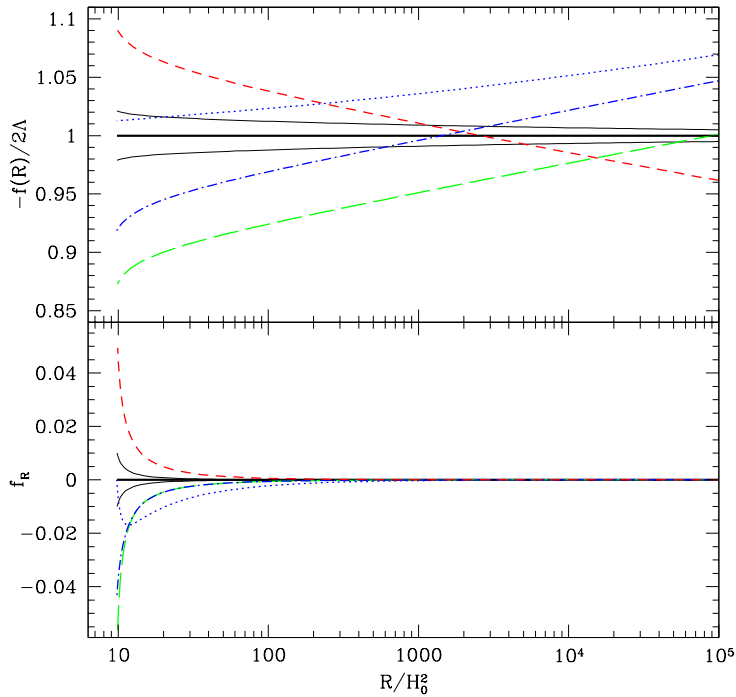}
\caption{$f(R)$ (top panel) and $f_R$ (bottom panel) vs $R$ obtained using the designer approach for several given expansion histories. The three solid black lines correspond to $w_{\rm{eff}}=-1$ with $A=0$ (f=-2$\Lambda$), $A<0$ ($-f/2\Lambda >1$, $f_R>0$) and $A>0$ ($-f/2\Lambda <1$, $f_R<0$). The red short-dashed line corresponds to $w_{\rm{eff}}=-1.01$ with $A=0$. The blue dot-dashed line is for $w_{\rm{eff}}=-0.99$ with $A=0$, while the blue dot line corresponds to $w_{\rm{eff}}=-0.99$ but with $A$ adjusted to make $|f_R|=10^{-6}$ today. The blue dot on the lower panel corresponds to $10\times f_R$ which makes it easier to see. Finally, the green long dash denotes a model with $A=0$ and a time varying $w_{\rm{eff}}(a)$ which crosses from $-0.99$ at early epochs to $-1.01$ for $z<1$.}
\label{fig:fofrw}
\end{figure}

The particular solution is proportional to $\rho_{\rm{eff}}$, whose absolute value is a decreasing (increasing) function of time for models with $w_{\rm{eff}}>-1$ ($w_{\rm{eff}}<-1$).  
In the special case of $w_{\rm{eff}}=-1$, the particular solution is a constant. In this case, all positive values of $A$ result in solutions satisfying Conditions~\ref{cond:1} and \ref{cond:3}, while all solutions with negative $A$ have $f_{RR}<0$ and $f_R>0$. One can also satisfy Condition~\ref{cond:chameleon} by choosing an appropriate $A$ in order to make $|f_R|$ as small as needed. Three sample solutions for $w_{\rm{eff}}=-1$ with $A=0$, $A<0$ and $A>0$ are shown with black solid lines in Fig.~\ref{fig:fofrw}. $A=0$ corresponds to the case of the cosmological constant, $f(R)=-2\Lambda$.

Let us next consider models with a constant $w_{\rm{eff}}\ne -1$. Particular solutions in the case of $w_{\rm{eff}}<-1$ are unacceptable because they lead to $f(R)$ that violate Conditions~\ref{cond:1} and \ref{cond:3}, as can be seen from the red short-dashed line in Fig.~\ref{fig:fofrw}. One could try to remedy the situation by choosing a sufficiently large positive $A$, to make $f_{RR}>0$. However, all such attempts lead to models that initially have $f_{RR}<0$ and then cross through $f_{RR}=0$ at some finite $R$. Such models would have instabilities and hence $w_{\rm{eff}}=const<-1$ is not allowed.

Models with $w_{\rm{eff}}=const>-1$ are generally better behaved, since the corresponding particular solution satisfies Conditions~\ref{cond:1} and \ref{cond:3} as can be seen by looking at the dot-dashed blue line in Fig.~\ref{fig:fofrw}. One does run into troubles, however, when trying to satisfy Condition~\ref{cond:chameleon} by making $|f_R|$ small today with a suitable choice of a negative $A$. Then the particular solution will still force $|f_R|$ to grow in time at high R, while the homogeneous part will bring it down to a small value today. Such $f_R$ will have a turning point and, hence, will violate both Conditions~\ref{cond:1} and \ref{cond:3}. This case is illustrated with a blue dotted line in Fig.~\ref{fig:fofrw}. However, if one chose to only study cosmological implications of $f(R)$ and ignore the galactic and solar constraints, then histories with $w_{\rm{eff}}=const>-1$ would generally be acceptable.

One gains more flexibility by allowing for a time-dependent equation of state. In particular, it is possible to construct $f(R)$ that have $w_{\rm{eff}}$ crossing $-1$ without violating Conditions~\ref{cond:1} and \ref{cond:3}. An example is shown in Fig.~\ref{fig:fofrw}, where the long-dashed green line corresponds to a model that has $w_{\rm{eff}}=-0.99$ at early times and transitions to $-1.01$ at redshift of $1$. 

Quite generally, Condition~\ref{cond:chameleon} discourages large deviations of $w_{\rm{eff}}$ from $-1$, since otherwise one must deal with turning points in $f_R$ at finite values of $R$. Hence, if we want to have reasonable physics on solar and galactic scales, we must have $w_{\rm{eff}}\approx -1$. As a consequence, if future data actually finds any deviations from $w_{\rm{eff}}=-1$, it would automatically rule all $f(R)$ models out. Only if galactic and solar tests are ignored, a richer variety of expansion histories can be considered.  For the numerical analysis in the rest of this Paper we will only use models with $w_{\rm{eff}}=-1$. 

Let us remark that, while such designer models appear extremely contrived, they are not necessarily more fine-tuned than quintessence dark energy models in terms of the number of free parameters. One could argue, in addition, that a universal order one coupling of the scalar field to matter is better motivated by high energy physics, where dilaton fields can appear naturally. Producing minimally coupled scalar fields requires additional fine-tuning. From the observational point of view, a quintessence dark energy with $w\rightarrow -1$ would be impossible to distinguish from $\Lambda$. The $f(R)$ models, on the other hand, predict a $\Lambda$CDM background expansion but detectably different dynamics of structure formation.

An additional potential problem with $f(R)$ models of ``dark energy'' was recently pointed out by Starobinsky \cite{Starobinsky:2007hu}. He observed that the perturbation $\delta R$ of the Ricci scalar in $f(R)$ models in the high curvature limit is a rapidly oscillating function of time, with an amplitude that can easily be larger than the average value of $R$. An oscillating background could lead to an overproduction of scalaron particles that would impact the Big Bang nucleosynthesis. Commenting on the quantum particle production would go beyond the scope of this Paper, although this is an important issue that should be further examined and clarified. At the classical level, however, it would seem that large rapid oscillations of $R$ do not necessarily imply a problem. $R$ is constructed from derivatives of the metric, so perturbations of the metric can remain tiny even when their derivatives are large when they undergo very rapid oscillations. In fact, as we show in the next Section, the metric potentials and the density perturbations are well behaved, despite $\delta R$ being large and oscillatory. In the equations for the metric, $\delta R$ always appears pre-multiplied by the small factor $f_{RR}$ which keeps the metric perturbations in the linear regime.

%%%%%%%%%%%%%%%%%%%%%%%%%%%%%%%%%%%%%%%%%%%%%%%
\section{Perturbation Theory in the Jordan frame}
\label{perttheory}
%%%%%%%%%%%%%%%%%%%%%%%%%%%%%%%%%%%%%%%%%%%%%%%

We consider scalar perturbations in the Newtonian gauge, with the line element given by
\be\label{metric}
ds^2= -a^2(1+2\Psi)d\tau^2+a^2(1-2\Phi)d\bar{x}^2 \ ,
\ee
where $\Psi=\Psi(\tau,\vec{x})$ and $\Phi=\Phi(\tau,\vec{x})$ are small time and space-dependent perturbations to the metric. In the rest of the paper all the perturbed quantities and the equations are presented in Fourier space, where the different $k$-modes evolve independently. Primes will indicate derivatives with respect to $\ln a$ and quantities such as $H$, $k$ and $f_{RR}$ are expressed in units of $H_0$ (or $H_0^2$) so that they are dimensionless.

For a general fluid component, the energy-momentum tensor has the following expansion at first order
\ba\label{en-mom-generic}
&&T^0_0=-\rho(1+\delta)\, ,\nonumber\\
&&T^0_i=-(\rho+P)v_i \equiv -\rho V_i\, ,\nonumber\\
&&T^i_J=(P+\delta P)\delta^i_j+\pi^i_j\, ,
\ea
where $\delta\equiv\delta\rho/\rho$ is the density contrast, $v$ is the velocity, $V \equiv (1+w)v$, $\delta P$ the pressure perturbation, and we define the anisotropic stress $\rho\Pi\equiv\l(\hat{k}^j\hat{k}_i-\f{1}{3}\delta^j_i\r)\pi^i_j$, where $\pi^i_j$ denotes the traceless component of the energy-momentum tensor.
In the Jordan frame, matter is minimally coupled and follows the geodesic of the usual metric $g_{\mu\nu}$. Thus, there is no explicit dependence of the matter Lagrangian ${\cal L}_{\rm m}[\chi_i,g_{\mu\nu}]$ on the $f(R)$ modifications, and the conservation equations for matter do not differ from the conservation equations of standard general relativity. We refer the reader to Appendix \ref{Einstein_linear} for the explicit expression of these equations, as well as for the full set of modified linear Einstein equations. Here we want to focus on the anisotropy and Poisson equations; the former relates the two Newtonian potentials $\Psi$ and $\Phi$, while the latter describes how the curvature potential $\Phi$ depends on the matter comoving density perturbation 
\be
\Delta\equiv\delta+3\f{aH}{k} V \ .
\ee 
For theories described by the action (\ref{jordanaction}), the {\it anisotropy} equation reads
\be\\ \label{anisotropy}
\Phi-\Psi=\f{9}{2}\f{a^2}{k^2}E_i\Pi_i-f_R\l(\Phi-\Psi\r)+f_{RR}\delta R \ ,
\ee
where $\delta R$ is the linear perturbation~(\ref{deltaR}) to the Ricci scalar $R$, repeated indices denote a sum over the matter fields, and we have defined $E_i$ in (\ref{E(x)}).
The {\it Poisson} equation, obtained combining equations
(\ref{time-time}) and (\ref{momentum}), is
\be\label{modified_Poisson}
\f{k^2}{a^2}\Phi=-\f{3}{2}E_i\Delta_i-\l[f_R\f{k^2}{a^2}\Phi-\f{1}{2}\f{k^2}{a^2}f_{RR}\delta R+\f{3}{2}H^2f_R'\l(\Psi+\Phi'\r)+\f{3}{2}HH' f_{RR}\delta R\r] \ .
\ee
Using equation (\ref{anisotropy}), and neglecting any anisotropic contribution from matter fields (i.e. setting $\Pi_i=0$), we obtain the following relation between the Newtonian potentials
\be\label{New_pot}
\Phi-\Psi=\f{f_{RR}}{F}\delta R
\ee
where we have defined $F\equiv1+f_R$.

In standard GR, once we neglect any matter shear, the anisotropy equation is simply a constraint, $\Psi=\Phi$, which reduces the number of independent perturbed variables. The Poisson equation is also an algebraic relation between the curvature perturbation $\Phi$ and the matter density perturbation $E_i\Delta_i$. 

On the contrary, in $f(R)$ gravity, eqs.~(\ref{modified_Poisson}) and (\ref{New_pot}) are dynamical.  Indeed, due to the higher order of the theory, the linear equations contain extra dynamics. In the specific case of the anisotropy equation, the extra dynamics is encoded in the  slip between the gravitational potentials. We will choose this slip as one of the perturbed variables to evolve. To render the equations more treatable numerically, and also to facilitate the physical interpretation of the results, we introduce new variables, functions of the Newtonian potentials $\Phi$ and $\Psi$. Namely, along with the perturbations of matter fields,  we choose to evolve the following variables 
\ba
\label{Phip}
&&\Phi_+\equiv\f{\Phi+\Psi}{2}\\
\label{chi}
&&\chi\equiv f_{RR}\delta R=F(\Phi-\Psi) \ .
\ea
Any non-zero value of the variable $\chi$ will signal a departure from standard GR. The variable $\Phi_+$ is the combination of potentials that affects propagation of light, leading to the Integrated Sachs-Wolfe (ISW) effect in the CMB and Weak Lensing (WL) of distant galaxies.

In terms of the variables (\ref{Phip})-(\ref{chi}), the anisotropy equation (\ref{New_pot}) becomes a constraint (which allows us to eliminate $\Psi$ in terms of $\Phi_+$ and $\chi$). We ignore the radiation component due to its relative unimportance for the late time dynamics of perturbations.
We can then obtain two coupled first order differential equations for $\chi$ and $\Phi_+$ from the momentum (\ref{momentum})  and the Poisson (\ref{modified_Poisson}) equations 
\ba\label{phiprime}
&&\Phi_+'=\f{3}{2}\f{aE_m V_m}{HkF}-\l(1+\f{1}{2}\f{F'}{F}\r)\Phi_++\f{3}{4}\f{F'}{F^2}\chi \\
\label{chiprime}
&&\chi'=-\f{2E_m\Delta_m}{H^2}\f{F}{F'}+\l(1+\f{F'}{F}-2\f{H'}{H}\f{F}{F'}\r)\chi-2F\Phi_+'-2F\l(1+\f{2}{3}\f{k^2}{a^2H^2}\f{F}{F'}\r)\Phi_+ \ .
\ea
\begin{figure}[t]
\includegraphics[width=86mm]{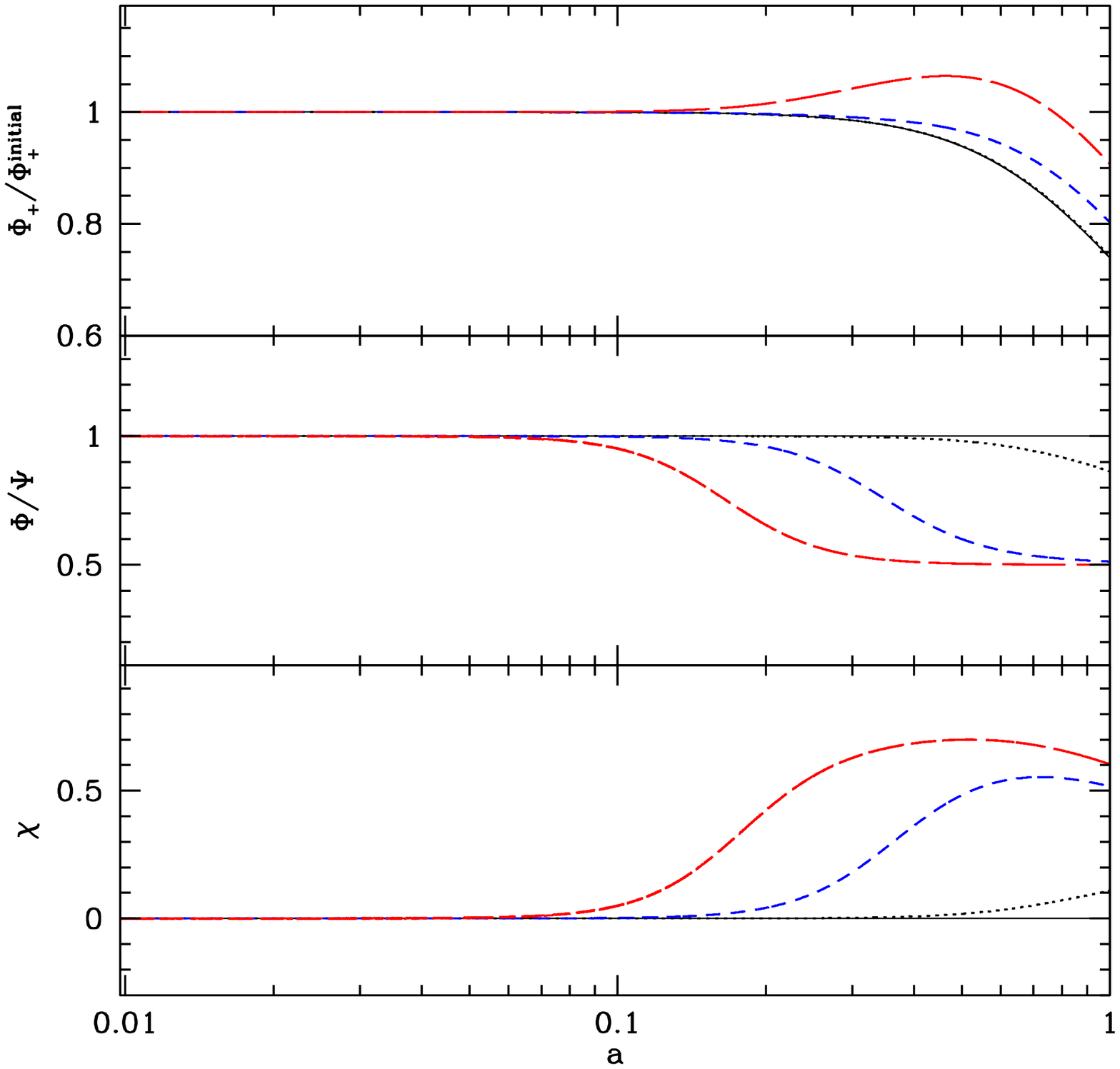}
\includegraphics[width=86mm]{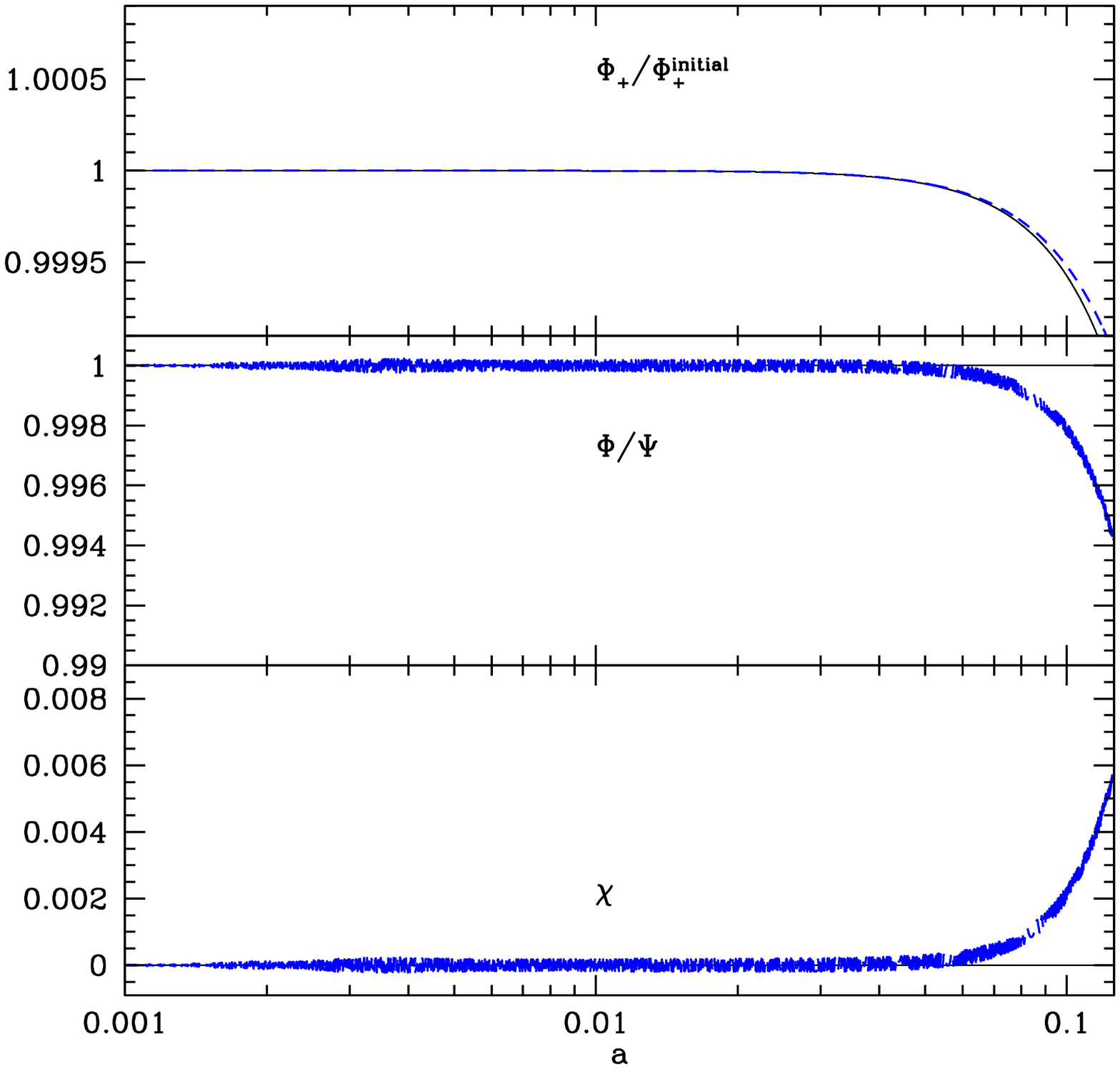}
\caption{These plots show the numerical solutions for the transfer function $\Phi_+/\Phi_+^{initial}$, the ratio of the potentials $\Phi/\Psi$ and the slip $\chi=f_{RR}\delta R$ as functions of the scale factor $a$  for different scales for an $f(R)$ model with $w_{\rm{eff}}=-1$ and $f_R^0=-10^{-4}$. The dotted line corresponds to the scale $k=0.01$h/Mpc,  the short-dashed line to  $k=0.1$h/Mpc and the long-dashed line to $k=0.5$h/Mpc. Finally, the solid line shows the scale-independent behavior in the $\Lambda$CDM case. The right panel shows the tiny oscillations in $\chi$ at early times, shown for $k=0.1$h/Mpc.}
\label{fig:pert}
\end{figure}
The numerical solutions of the above equations are shown in Fig.~\ref{fig:pert} for an $f(R)$ model with $w_{\rm{eff}}=-1$ and $f_R^0\equiv f_R(z=0)$ equal to $-10^{-4}$. Such a value of $f_R^0$ is outside the bounds of Condition~\ref{cond:chameleon} of Subsection~\ref{viable}. However, we choose to use this model in all of our plots because the pattern of $f(R)$ modifications can be easily seen ``by eye'', while all qualitative features of the perturbation dynamics are still the same as in models with $|f_R^0|<10^{-6}$, that are more likely to satisfy Condition~\ref{cond:chameleon}. We start evolving the equations at $z=1000$, when the deviations from GR are small, and set $\Phi_+=-1$ and $\chi=0$ at the initial time. We then use the standard GR relations  
\be
v_m={2k \over 3aH}\Phi_+ \ \ , \ \Delta_m = -{2k^2 \over 3 a^2H^2}\Phi_+
\ee
to set the initial conditions for the matter perturbations $\Delta_m$ and $v_m$.

At early times, when the effect of $f(R)$ on the background expansion is negligible, and when the wavelength corresponding to $k$ is larger than $\lambda_C$ (but still inside the horizon), the potential stays constant as one would expect in GR. As soon as the mode enters inside the Compton radius, the additional attractive force mediated by the scalaron starts to enhance the growth. The potential eventually starts to decay when the universe begins to accelerate. The plot of $\Phi_+$ in the left panel of Fig.~\ref{fig:pert} clearly shows these two trends -- the enhancement due to the ``fifth force'' and decay due to an accelerating background. In the right panel of Fig.~\ref{fig:pert} we show a magnified plot of the evolution of the same quantities at early times. One can see that $\chi$ oscillates rapidly, but the amplitude is very small. These oscillations correspond to very large oscillations of $\delta R=\chi/f_{RR}$ (also pointed out in \cite{Starobinsky:2007hu}), but they do not have a noticeable effect on the evolution of the metric perturbations.

To better understand the asymptotic values of $\chi$ and $\Phi/\Psi$ in Fig.\ref{fig:pert},  we analytically examine the evolution of perturbations in the Jordan and Einstein frames in the sub-horizon regime. We then interpret the modified dynamics of perturbations in $f(R)$ in terms of an effective dark energy fluid with a non-zero shear.

%%%%%%%%%%%%%%%%%%%%%%%%%%%%%%%%%%%%%%%%%%%%%%%
\subsection{The behavior of perturbations on sub-horizon scales}
\label{sub-horizon}
We are interested in the behavior of matter and metric perturbations for modes that are inside the horizon well after the radiation-matter equality. Therefore, in the following analytical study we neglect radiation and deal just with cold dark matter (CDM) characterized by $w=0$. It is a well known result of standard GR that the CDM  density perturbations grow linearly with the scale factor on sub-horizon scales during matter domination, while the potentials $\Phi$ and $\Psi$ stay constant. When the cosmological constant begins to dominate, the growth of structures slows down and the Newtonian potentials start decaying, resulting in a late-time Integrated Sachs-Wolfe (ISW) signal which contributes to the low multipoles of the CMB. These results can be easily derived by considering the sub-horizon, $k/a\gg H$, version of the evolution equations. 

In $f(R)$ theories, however, there is an extra length scale associated with the additional scalar dynamical degree of freedom -- the Compton wavelength $\lambda_C$~(\ref{lambda_C}) of the scalaron. This scale separates two regimes of sub-horizon gravitational dynamics~\cite{Saw-Hu:2007,Tsujikawa:2007}, so the sub-horizon approximation must be taken with more care. On scales $\lambda\gg\lambda_C$, the scalar field is massive, the ``fifth force'' is exponentially suppressed, thus deviations from GR are negligible. However, on scales below the Compton wavelength, the scalaron is light and deviations are significant. The relations between $\Phi$ and $\Psi$, and between them and the matter density contrast will be different below the Compton scale and that affects the growth rate of structures. Different approaches to the study of the growth of structures have already been employed in the literature~\cite{Zhang:2005vt,Song:2006ej,Am-Tsujikawa:2007}. It is instructive to compare these approaches. We do it in the following subsections, where we also introduce a new method in the variables $\chi$ and $\Phi_+$ , as well as perform a parallel analysis in the Einstein frame.

%%%%%%%%%%%%%%%%%

\subsubsection{In Jordan frame}
On sub-horizon scales, the equations for $\Phi_+$ and $\chi$ reduce to
\ba
\label{Poisson-sub-hor}
&&\frac{k^2}{a^2}\Phi_+\simeq-\frac{3}{2}\frac{E_m\delta_m}{F}\\
\label{chi-Phi+}
&&\chi\simeq-\f{2f_{RR}k^2}{3f_{RR}k^2+a^2F}F\Phi_+\ ,
\ea
where we have used $\Delta\simeq\delta$  on sub-horizon scales. From (\ref{chi-Phi+}), it is clear that the behavior of $\chi$  is characterized by a transition scale set by the Compton wavelength of the scalaron $\lambda_C$ (\ref{lambda_C}).

It is useful to introduce a parameter $Q$, approximately defined as the squared ratio of the Compton wavelength to the physical wavelength of a mode\footnote{A similar parameter $Q$ was first introduced in~\cite{Zhang:2005vt}. Here we choose to have it positive and to correspond to the ratio of the wavelengths.}
\be\label{Q}
Q\equiv3\f{k^2}{a^2}\f{f_{RR}}{F} \approx\l(\f{\lambda_C}{\lambda}\r)^2\ .
\ee
Then we can rewrite eq.~(\ref{chi-Phi+}) as 
\be\label{chi-Phi+-sub}
\chi\simeq-\f{2}{3}\f{Q}{1+Q}F\Phi_+ \ .
\ee
For scales $\lambda\gg\lambda_C$, the parameter $Q\rightarrow0$ and the variable $\chi$ is negligibly small. However, for scales below $\lambda_C$, $Q\gg 1$ and $\chi$ becomes of the same order as $\Phi$ and $\Psi$, asymptotically approaching $\chi\simeq-2/3 F\Phi_+$. This can also be seen from the plots in Fig.~\ref{fig:pert}.

Next, we want to understand the impact of the relation (\ref{chi-Phi+-sub}) on the evolution of the gravitational potentials and the growth of density perturbations. We choose to do it in the variables $\Phi$ and $\Psi$ to render the comparison with other literature easier. From eq.(\ref{chi-Phi+-sub}) it follows that
\be\label{Psi-Phi-sub-Q}
\Psi\simeq\f{3+4Q}{3+2Q}\Phi \ .
\ee
\begin{figure}[t]
\includegraphics[width=150mm]{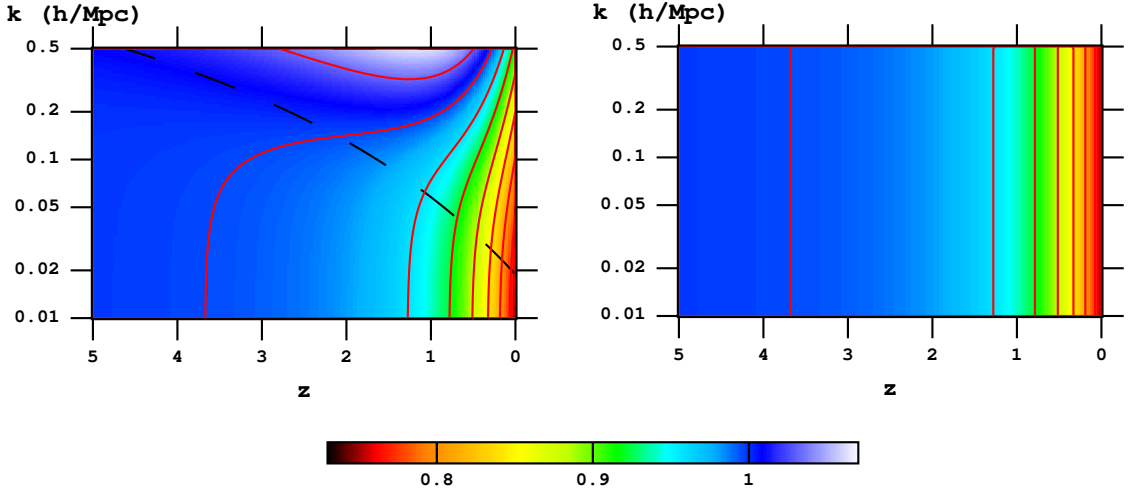}
\caption{The evolution of the growth factor for the CDM $[\Delta_m(k,a)/a]/[\Delta_m(k,a_i)/a_i]$ as a function of redshift $z$ and scale $k$. The left panel corresponds to an $f(R)$ model with $w_{\rm{eff}}=-1$ and $f_R^0=-10^{-4}$. In the right panel we show the corresponding $\Lambda$CDM pattern for comparison. One can see the scale-dependent behavior of the growth factor in $f(R)$ as opposed to the scale-independence of the $\Lambda$CDM case. The dashed line crossing diagonally on the left plot corresponds to the Compton transition wavelength given by $Q=1$ (\ref{Q}). Deviations from $\Lambda$CDM become important on scales below $\lambda_C$. For modes with $\lambda<\lambda_C$ during matter domination, there is an enhancement in the growth due to the "fifth force" introduced by the modifications to GR. Eventually, the universe starts accelerating and the growth slows down. However, in comparison to the $\Lambda$CDM case, such slowing is delayed in a scale dependent way.}
\label{fig:growth}
\end{figure}
On scales $\lambda\gg\lambda_C$, $Q\rightarrow 0$, and the standard relation $\Psi\simeq\Phi$ still holds. However, on scales below $\lambda_C$, the relation between the metric potentials becomes $\Psi\simeq2\Phi$. 

In order to see how the growth of structures is affected by $f(R)$ modifications, let us combine eqs.(\ref{mattereom-generic1})-(\ref{mattereom-generic2}) into a second order differential equation for the matter density contrast. On sub-horizon scales, considering only CDM, the equation is
\be\label{mattersecond}
\delta_m''+\l(1+\f{H'}{H}\r)\delta_m'+\f{k^2}{a^2H^2}\Psi=0 \ .
\ee
Combining the sub-horizon version of the Poisson equation (\ref{modified_Poisson}), with (\ref{Psi-Phi-sub-Q}), we can derive an expression for the potential $\Psi$ in terms of the matter density perturbation
\be\label{Psi-sub-matter}
\f{k^2}{a^2}\Psi\simeq-\f{3}{2}\f{1}{F}\f{1}{3}\f{3+4Q}{1+Q}E_m\delta_m \ .
\ee
One can interpret the modification as the Newton's constant being rescaled in a time, scale and model dependent way, inducing a scale (and model) dependence in the growth of structures. In particular, for a mode inside the horizon but well above the Compton wavelength, $Q\ll1$, the only effect of the $f(R) $ modifications to the growth of structure is a $1/F$ rescaling of the constant $G$. But when the mode is well inside the Compton scale, the Newton's constant is rescaled by a factor $4/(3F)$. Hence, we have two regimes separated by the scalaron Compton wavelength:
\ba\label{2regimes}
&&\Psi\simeq\Phi\,\,,\chi\simeq0\,\,,\,\, G_{\rm{eff}}\simeq\f{G}{F}\,\,,\,\,\,\,\,\,\,\,\,\,\,\,\,for\,\,\,\,\,\lambda\gg\lambda_C\nonumber\\
&&\Psi\simeq2\Phi\,\,,\chi\simeq-\f{2}{3}F\Phi_+\,\,,\,\,
G_{\rm{eff}}\simeq\f{4}{3}\f{G}{F}\,\,,\,\,\,\,\,\,\,for\,\,\,\,\,\lambda\ll\lambda_C
\ea
This effect can be clearly seen in Fig.~\ref{fig:growth}. On scales smaller than the scalaron Compton wavelength, the modifications introduce a fifth force which would enhance the growth of structures. The rate of growth will depend on the balance between the fifth force and the background expansion. For modes that cross below $\lambda_C$ during matter domination, the effect of modifications is maximized, as the potential $\Phi_+$ grows in the absence of background acceleration. When, however, the background expansion starts accelerating, it compensates for the enhancement due to the fifth force. Eventually, potentials start to decay, but at a lesser rate than in the $\Lambda$CDM model. 
\begin{figure}[t]
\includegraphics[width=150mm]{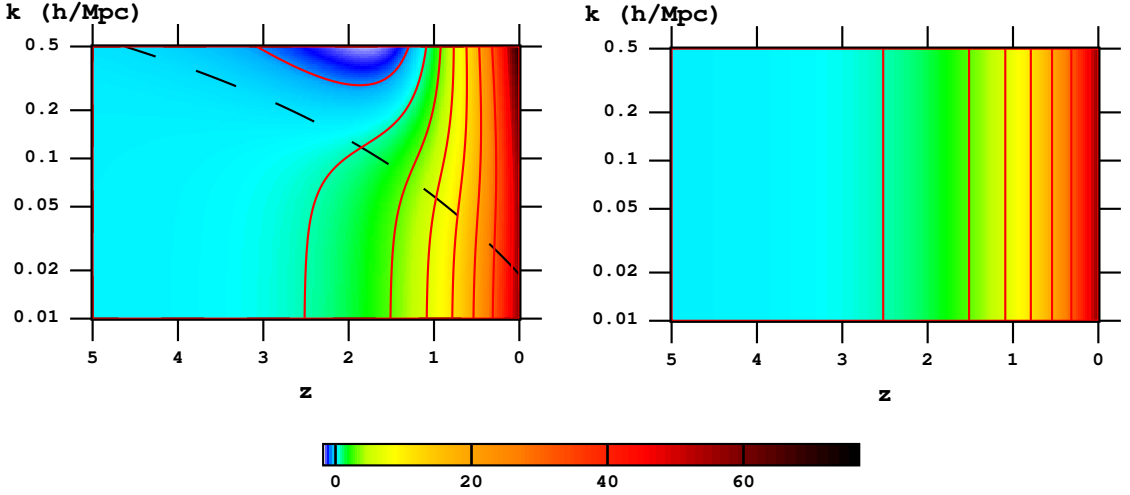}
\caption{The function probed by the cross-correlation of large scale structure with ISW, $\Delta_m\cdot d\Phi_+/dz$, as a function of scale $k$ and redshift $z$. The left panel corresponds to an $f(R)$ model with $w_{\rm{eff}}=-1$ and $f_R^0=-10^{-4}$ and shows a characteristic scale-dependent pattern. The right panel corresponds to $\Lambda$CDM. The dashed line crossing through the left panel corresponds to $Q=1$ (\ref{Q}), i.e. it  corresponds to the Compton wavelength of the scalaron  $\lambda_C$ (\ref{lambda_C}). For scales $\lambda<\lambda_C$, during matter domination, one can clearly notice the effect of the "fifth force" which suppresses the cross correlation and can actually make the correlation negative. Therefore, a negative cross correlation signal at early redshifts (corresponding to matter era), is a signature of $f(R)$. The acceleration of the background will eventually contrast the "fifth force" and lead to a positive cross-correlation.}
\label{fig:isw}
\end{figure}

A characteristic observational signature of $f(R)$ models would be an ISW effect during the matter era if one could correlate the distribution of large scale structure at $z>2$ with the CMB. In fact, such a correlation would be {\it negative}, since potentials would be growing, and not being constant as in the usual case. This is clearly seen in Fig.~\ref{fig:isw}. One must be realistic, however, and keep in mind that large statistical errors associated with the ISW measurements, and the smallness of the effect expected in viable $f(R)$ models, will likely render the ISW based tests of $f(R)$ useless. Unless, of course, missions such as LSST \cite{LSST} and SKA \cite{SKA}, which will be able to probe structures at redshifts  $z \lesssim 3$ and $z>3$ respectively, find a statistically significant positive LSS-ISW correlation signal at high redshifts, which would effectively rule out $f(R)$ (and $\Lambda$CDM).

From eq.(\ref{Poisson-sub-hor}), we notice that the potential $\Phi_+$ evolves in time as $\Delta_m/a$, up to a time-dependent factor $F$ which for viable models is practically equal to $1$. Such behavior is analogous to that in $\Lambda$CDM when $F\simeq 1$. Of course, in the case of $f(R)$ the evolution of $\Delta_m$ is modified, yet it is an interesting feature of $f(R)$ theories that this relation between $\Phi_+$ and $\Delta_m$ is approximately preserved (up to a factor $F\simeq 1$) on sub-horizon scales. A similar result also holds for the Modified Source Gravity model~\cite{Sean}. In the Dvali-Gabadadze-Porrati model \cite{Dvali:2000hr}, $\Phi_+$ evolves exactly like $\Delta_m/a$ \cite{KoyamaDGP} without any time-dependent prefactors.

\begin{figure}[t]
\includegraphics[width=80mm]{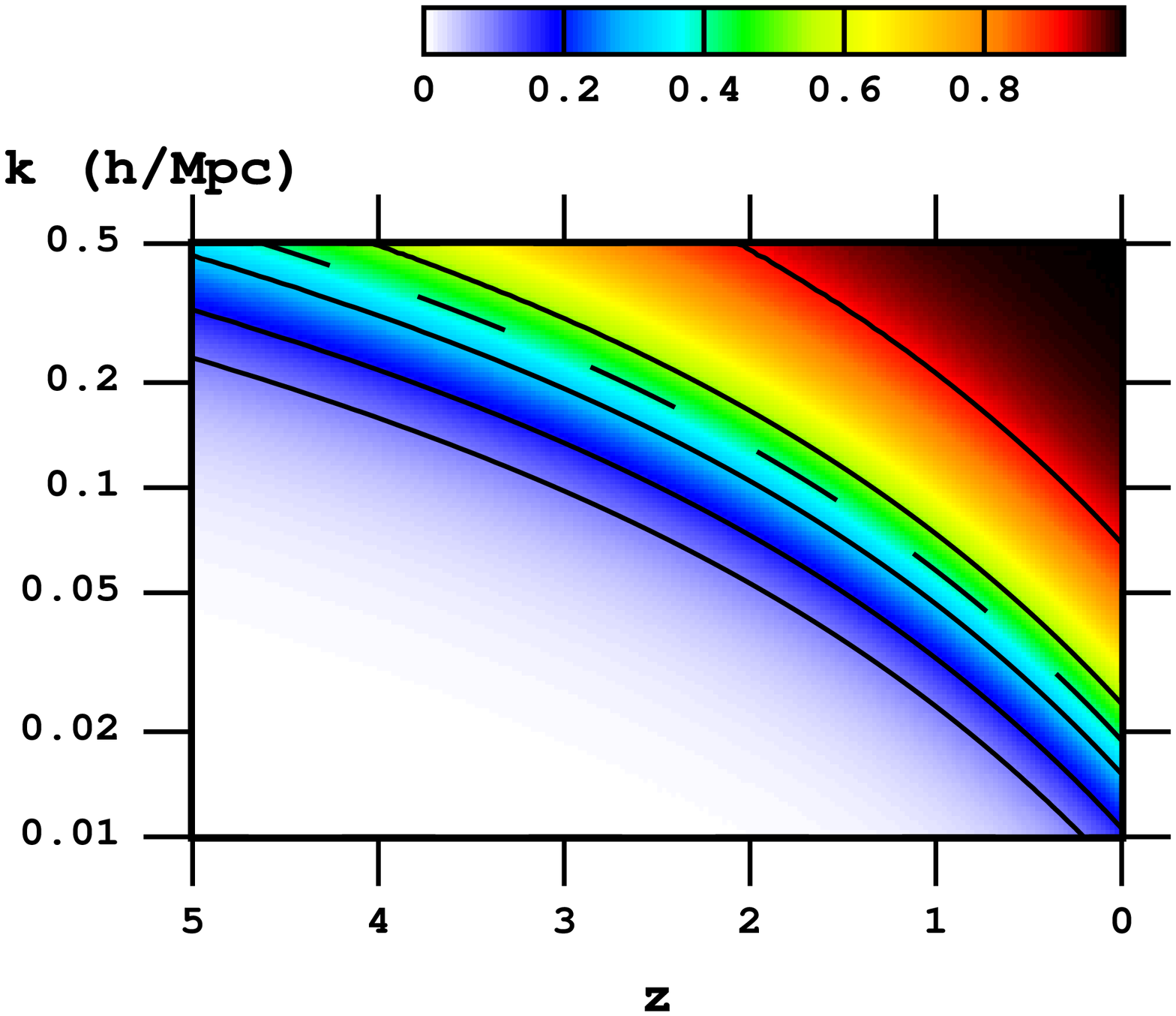}
\includegraphics[width=80mm]{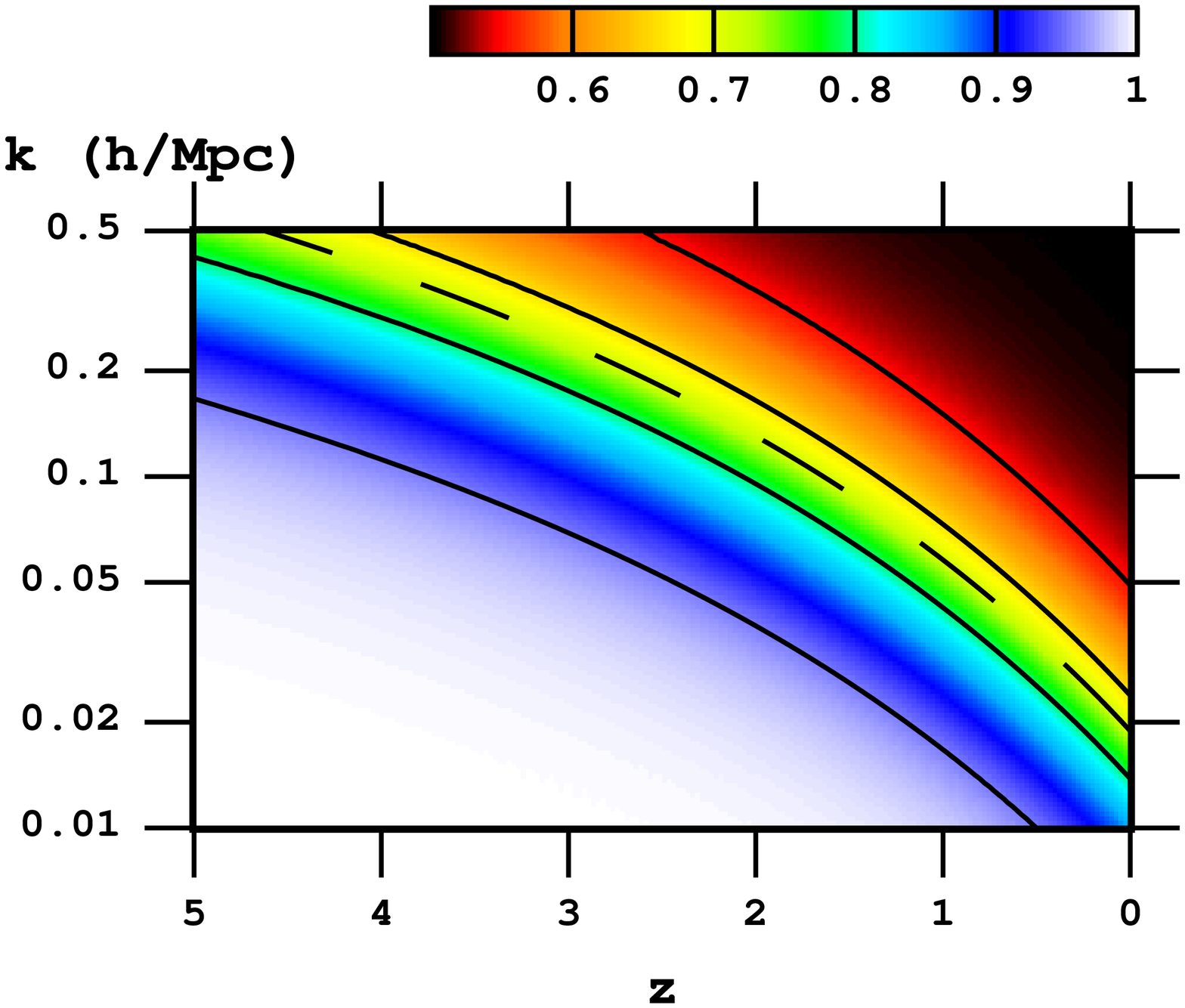}
\caption{The parameters $\bo=(\Psi-\Phi)/\Phi$ (left) and $\eta=\Phi/\Psi$ (right) as functions of scale $k$ and redshift $z$ for an $f(R)$ model with $w_{\rm{eff}}=-1$ and $f_R^0=-10^{-4}$. The long-dashed line corresponds to the Compton transition scale $Q=1$ (\ref{Q}). One can notice the time- and scale-dependent pattern. The parameter $\bo$ evolves from $\bo\simeq0$ on scales above $\lambda_C$ (\ref{lambda_C}), to $\bo\simeq1$ on scales $\lambda\ll\lambda_C$. Analogously, $\eta$ evolves from $\eta\simeq1/2$ for $\lambda\gg\lambda_C$ to $\eta\simeq1$ for scales below $\lambda_C$.}
\label{fig:omega-eta}
\end{figure}

The change in the relation between the Newtonian potentials $\Phi$ and $\Psi$ has been the focus of some recent literature~\cite{Caldwell,Bean-Dod}. In~\cite{Caldwell}, a parameter $\bo$ was introduced to parametrize the slip between the potentials. In the case of $f(R)$ theories, $\bo$ is given by the following expression
\be\label{omegabar_fR}
\bo\equiv\f{\Psi-\Phi}{\Phi}=-\f{f_{RR}\delta R}{F\Phi} \ .
\ee
Its evolution as a function of redshift and scale is shown in Fig.~\ref{fig:omega-eta}.
On sub-horizon scales, we can approximately write (\ref{omegabar_fR}) as
\be\label{omega}
\bo\simeq\f{2Q}{2Q+3} \ .
\ee
We notice that $\bo$ is in general a model, time and scale dependent parameter; on sub-horizon scales, it evolves from $\bo\simeq0$ on scales $\lambda\gg\lambda_C$, to $\bo\simeq 1$ on $\lambda\ll\lambda_C$.\\

In~\cite{Bean-Dod}, an analogous parameter was introduced, $\eta\equiv\Phi/\Psi$ \footnote{Note the different sign convention for our $\Phi$ compared to that in \cite{Bean-Dod}.}. On sub-horizon scales, this time, scale and model dependent parameter assumes the following expression
\be\label{eta_fR}
\eta\simeq\f{3+2Q}{3+4Q}
\ee
Hence it evolves from $\eta\simeq1$ on scales above the Compton one, to the value $\eta\simeq1/2$ on scales smaller than the Compton scale, as can be seen in Fig.~\ref{fig:omega-eta}.

%%%%%%%%%%%%%%%%

\subsubsection{In Einstein frame}

It is instructive to derive the above results working in the Einstein frame. In this frame, the modifications are described by a scalar field $\phi$ which couples to matter, as described in Appendix \ref{conformal}. The perturbation to the scalar field, $\delta\phi$, is related to the variable $\chi$ via the conformal transformation from the Einstein to the Jordan frame, $\chi/F=\beta\kappa \delta\phi$. The behavior of perturbations for a coupled dark energy field has been studied in details in~\cite{Amendola:2003}, and for the case of Chameleon fields in~\cite{Bruck}; here we adapt it to the specific case of $f(R)$ theories. The full linear evolution equations for the perturbation to the scalar field and for the CDM density contrast are included in Appendix \ref{conformal}. In this subsection we focus on the sub-horizon regime.

We notice from (\ref{cdmsecond-Einstein}), that the effective potential acting on CDM is $\tilde{\Psi}- \beta\kappa\delta\phi/2$, where the tilde  indicates the Einstein frame quantities as in Appendix \ref{conformal}. On sub-horizon scales, the equation for the scalar field (\ref{phisecond}) reduces to  the non-linear Klein-Gordon equation for a massive field, namely
\be\label{phiYuk}
\delta\phi\simeq\f{3}{2}\f{\beta}{\kappa}\f{\tilde{a}^2\tilde{E}_m\tilde{\delta}_m}{k^2+\tilde{a}^2m_{\phi}^2}
\ee
Given this relation between $\delta\phi$ and $\tilde{\delta}_m$, on sub-horizon scales ($aH/k\ll1$) the dominant contribution to the Poisson equation comes from the CDM, and  the effective potential acting on a CDM particle is 
\be\label{effEinstein}
%\tilde{\Psi}_{cdm}=
\tilde{\Psi}-\f{\beta\kappa\delta\phi}{2}\simeq-\f{3}{2}
\f{\tilde{a}^2}{k^2}\tilde{E}_m\tilde{\delta}_m\l[1+\f{1}{3}\f{k^2}{k^2+\tilde{a}^2m_{\phi}^2}\r]
\ee
Therefore the effect of the scalar field is to introduce a Yukawa type gravitational potential, with a characteristic scale $\lambda_{\phi}=2\pi/m_{\phi}$. The mass of the scalar field is uniquely determined as a  function of $f(R)$
\be\label{mass-scalar}
m_{\phi}^2\equiv V_{,\phi\phi}=\f{1}{3}\l[\f{1}{f_{RR}}+\f{FR-4(f+R)}{F^2}\r]\simeq \f{1}{3f_{RR}} \ .
\ee
Therefore the characteristic radius of the Yukawa type potential is $\lambda_{\phi}\simeq (3f_{RR})^{1/2}$. Let us map the entire equation (\ref{effEinstein}) to the Jordan frame. We use the mapping recipe described in~\cite{Bean:2006up}; in particular, the relation between the spatial curvature perturbations gives $\tilde{\Psi}=\Psi+\beta\kappa\delta\phi/2$ and for the energy density of CDM we have $\tilde{\delta}_m=\delta_m-2\beta\kappa\delta\phi$. Finally, the scale factors in the two frames are related via $\tilde{a}^2=e^{\beta\kappa\phi}a^2=Fa^2$. Applying all this to (\ref{effEinstein}), we reproduce eq.(\ref{Psi-sub-matter}). In particular we see that the physical scale $\lambda_{\phi}$, maps to the Compton scale $\lambda_{\phi}\rightarrow(3f_{RR}/F)^{1/2}\approx\lambda_C$, and that the value of the Newton's constant is rescaled as in (\ref{2regimes}). 

%%%%%%%%%%%%%%%%%%%%%

\subsubsection{Effective dark fluid interpretation}

\begin{figure}[t]
\includegraphics[width=85mm]{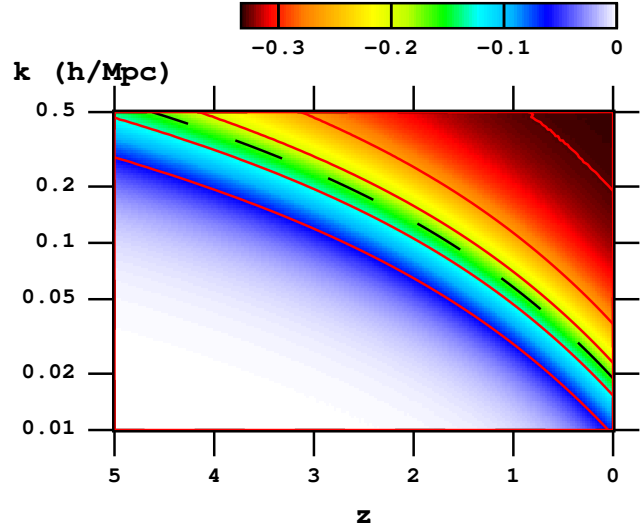}
\caption{The ratio $E_{\rm{eff}}\Delta_{\rm{eff}}/(E_m \Delta_m)$ as a function of $k$ and $z$. At early times and large scales the ratio is very small. After a given $k$-mode crosses into the Compton radius, the ratio monotonically evolves towards $-1/3$. The dash line corresponds to $Q=1$ (\ref{Q}).}
\label{fig:deff}
\end{figure}

While working in the Jordan frame, we can consider all the modifications to the Einstein tensor $G_{\mu\nu}$ as components of the {\it effective} energy-momentum tensor $T_{\mu\nu}^{\rm{eff}}$ (\ref{G_eff}). The fluid described by this tensor will not be an independent component, but a function of the metric. We can read off the linearly perturbed components of $T_{\mu\nu}^{\rm{eff}}$ directly from the Einstein equations in Appendix \ref{Einstein_linear}. Given the covariant conservation of the effective energy-momentum tensor (\ref{G_eff}), we need just two quantities to describe the fluid perturbations. We will choose the energy density contrast in the rest frame of the fluid, $\Delta_{\rm{eff}}$, and either the anisotropic stress, $\Pi_{\rm{eff}}$, or the sound speed, $c_{\rm{eff}}^2$, depending on the context.
 
The anisotropic stress associated to the effective dark fluid is
 \be\label{shear_eff}
E_{\rm{eff}}\Pi_{\rm{eff}}(k,a)\equiv \f{2}{9}\f{k^2}{a^2}\f{f_{RR}\delta R}{F}=\f{2}{9}\f{k^2}{a^2}\f{\chi}{F} \ ,
\ee
while  the effective comoving density perturbation is
\be\label{comdelta_eff}
E_{\rm{eff}}\Delta_{\rm{eff}}\equiv\f{2H^2}{3}\l[f_R\f{k^2}{a^2H^2}\Phi-\f{1}{2}\f{k^2}{a^2H^2}f_{RR}\delta R+\f{3}{2}f_R'\l(\Psi+\Phi'\r)+\f{3}{2}\f{H'}{H}f_{RR}\delta R\r]
\ee
Equivalently, we can derive the sound speed, defined as
\be\label{eff_soundspeed}
c_{\rm{eff}}^2\equiv\f{\delta P_{\rm{eff}}^{(c)}}{\rho_{\rm{eff}}\Delta_{\rm{eff}}} \ .
\ee

We can now interpret the results of the previous subsections in terms of this effective fluid. On sub-horizon scales ($k/a\gg H$), the effective dark fluid is characterized by the following density perturbation
 \be\label{eff-density-subhor}
E_{\rm{eff}}\Delta_{\rm{eff}}\simeq \f{1}{3}\f{k^2}{a^2}\f{Q(3F-2)+(F-1)}{1+3Q}\l(\Phi+\Psi\r)
\ee
and sound speed 
\be\label{eff-soundspeed-subhor}
c^2_{\rm{eff}}\simeq\f{2}{3}\f{Q}{(1-F)(1+2Q)+QF} \ .
\ee
The effective shear (\ref{shear_eff}) is proportional to $\chi$, therefore its behavior on sub-horizon scales is well described by (\ref{chi-Phi+-sub}). Above the scalaron Compton wavelength, the sound speed is small and positive, $c^2_{\rm{eff}}\simeq 0^+$, i.e. the effective fluid behaves as an extra sub-dominant clustering component with negligible shear. Eventually, the effective shear $\Pi_{\rm{eff}}\simeq {\chi}$ starts growing and tends to a value comparable to $\Phi_+$ on scales well below $\lambda_C$. The sound speed tends to the negative value $c^2_{\rm{eff}}\simeq-2/[3(3F-2)]\simeq-2/3$ for $\lambda\ll\lambda_C$. A negative value for the sound speed raises concerns about the stability of perturbations in the dark component. It would certainly lead to instabilities if there was no anisotropic stress associated with the effective fluid. However, in an imperfect fluid with anisotropic stress, one can achieve stability even with a negative value of the sound speed. This can be seen easily by combining the evolution equations (\ref{mattereom-generic1})-(\ref{mattereom-generic2}) for a generic fluid into a second order differential equation for $\delta$. The equation reads
\ba\label{second-cdm}
&&\delta''+3(1+w)\Phi''+\l[(1-3w)+\f{w'+\l(1+\f{H'}{H}\r)}{1+w}\r]\delta'+3w'\Phi'+\nonumber\\
&&+\l[3(c^2_s-w)\l(1-3w+\f{w'+\l(1+\f{H'}{H}\r)}{1+w}\r)+\f{k^2}{a^2H^2}\l(c_s^2-
\f{\Pi}{\delta}\r)\r]\delta+\f{k^2}{a^2H^2}(1+w)\Psi=0
\ea
The ratio of the anisotropic stress to the energy-density perturbation can be thought of as an anisotropic sound speed, $c_{an}^2\equiv \Pi/\delta$, which, combined with the usual sound speed, can stabilize the perturbations~\footnote{It is important to remember that the ratio of the pressure perturbation to the energy density contrast does not necessarily represent the physical velocity with which perturbations in the fluid propagate. Perhaps the term isotropic and anisotropic "stiffness"~\cite{Bashinsky} might be more appropriate, but we choose to employ the more commonly used term "sound speed".}. The stability condition in the presence of shear is then generalized to~\cite{Caldwell}
\be\label{eff-stability}
c^2_s-c_{an}^2 \geq  0
\ee
which is satisfied by the effective dark fluid.

In Fig.~\ref{fig:deff} we plot the ratio of the effective dark energy density perturbation to that of the CDM as a function of $k$ and $z$. At early times and large scales, the ratio is very small and $\Delta_{\rm{eff}}$ is oscillating around zero with a negligibly small amplitude. As soon as a given $k$-mode crosses into the Compton radius, the ratio begins to evolve monotonically towards $-1/3$. This asymptotic value can also be obtained analytically, by combining eq.(\ref{Poisson-sub-hor}) with eq.(\ref{eff-density-subhor}) (with $\Phi_+=(\Phi+\Psi)/2$) to get
\be\label{deltaeff_deltam}
\f{E_{\rm{eff}}\Delta_{\rm{eff}}}{E_m\Delta_m}\simeq-\f{Q(3F-2)+(F-1)}{1+3Q}
\ee
In the limit of $Q\gg1$, corresponding to scales well within the Compton radius,  the ratio (\ref{deltaeff_deltam}) tends indeed to $-1/3$.
This value is expected because it corresponds to the change of $G_{\rm{eff}}$ from $G/F$ to $4G/(3F)$ (\ref{2regimes}).
%%%%%%%%%%%%%%%%%%%%%%%%%%%%%%%%%%%%%%%%%%%%%%%
\section{Observational outlook and conclusions}

\label{conclusions}

We have studied the growth of structures in viable $f(R)$ theories. In order to construct $f(R)$ models with specific expansion histories we have used the "designer" procedure introduced in~\cite{Song:2006ej}, and generalized it to include radiation and a varying effective equation of state $w_{\rm{eff}}(a)$. We have shown that viable models, which satisfy all of the conditions reviewed in Subsection~\ref{viable}, must closely mimic the $\Lambda$CDM expansion history.  In fact, if any departure from $w_{\rm{eff}}=-1$ was observed, it would automatically rule out $f(R)$ as a fundamental theory aimed at explaining the cosmic acceleration. One could, in principle, have a viable $f(R)$ cosmology with $w_{\rm{eff}}\ne -1$ which would not, however, satisfy any of the local tests of gravity. 

The degeneracy with $\Lambda$CDM is broken when one considers the linear perturbations. $f(R)$ models predict a characteristic scale-dependent growth of large scale structures which may be observationally detectable. To illustrate this pattern, throughout the paper we employed the $f(R)$ model with $w_{\rm{eff}}=-1$ and $f_R^0=-10^{-4}$. Models with this value of $f_R^0$ would likely fail the solar and galactic constraints which, according to the analysis in \cite{Saw-Hu:2007}, require $|f_R^0| \lesssim 10^{-6}$. However, the model we used allowed us to demonstrate the qualitative features of the growth pattern better. The pattern in the model with $f_R^0=-10^{-6}$ would be similar, but with the modifications becoming graphically visible at correspondingly smaller scales. Observational signatures of models with $|f_R^0| \lesssim 10^{-6}$ would be quite subtle, at a level of a few percent difference in the growth factor \cite{Saw-Hu:2007}. Nevertheless, such a level of accuracy should still be within the reach of future weak lensing surveys.

We have solved the exact equations for linear perturbations in the Jordan frame numerically and then confirmed our results analytically in the Jordan and Einstein frames using the sub-horizon approximation. We observe a characteristic scale-dependent pattern in the growth of structures. The Compton wavelength of the  scalaron (\ref{lambda_C}) introduces a scale which separates two regimes of sub-horizon gravitational dynamics during which gravity behaves differently. On scales $\lambda\gg\lambda_C$, the scalaron is massive and the ``fifth force''  is exponentially suppressed, thus deviations from GR are negligible. However, on scales inside the Compton radius, the scalaron is light and deviations are significant. The relations between $\Phi$ and $\Psi$, and the relation between them and the matter density contrast, will be different below the Compton scale and that affects the growth rate of structures. In particular, on scales below $\lambda_C$, the modifications introduce a slip between these potentials, leading to $\Psi\simeq 2\Phi$.  The rate of growth of structures depends on the balance between the ``fifth force'' and the background acceleration. For modes that cross below $\lambda_C$ during matter domination the effect of modifications is maximized as the potential $\Phi_+$ grows in the absence of background acceleration. When, however, the background starts accelerating, the potentials begin to decay but at a lesser rate than in the $\Lambda$CDM model. Therefore, a characteristic signature of $f(R)$ theories would be a non-zero ISW effect during the matter era. One should note, however, that the ISW measurements are plagued by large statistical errors (because the ISW effect is only a part of the total CMB anisotropy) and cannot provide the percent level accuracy needed to test viable $f(R)$ models. Weak lensing studies, on the other hand, are counted on to eventually provide highly accurate $3$D maps of the gravitational potential $\Phi_+$. While a comprehensive quantitative analysis of the level at which they will be able to constrain $f_R^0$ is still lacking, our preliminary results, and those of \cite{Saw-Hu:2007,Schimd:2004nq,Amendola:2007rr} indicate that viable $f(R)$ models can still lead to observable differences from $\Lambda$CDM.

To conclude, we note that the $f(R)$ pattern of structure growth would be similar to that in Chameleon models \cite{Bruck} with a universal coupling of the chameleon to all matter fields. These models are somewhat less severely constrained than the $f(R)$ models because of the additional freedom in choosing the value of the coupling.

%%%%%%%%%%%%%%%%%%%%%%%%%%%%%%%%%%%%%%%%%%%%%%%

\acknowledgments

We thank A.~Frolov, M.~Kunz, M.~Trodden, C.~van de Bruck and S.~Winitzki for helpful discussions during the course of this work. The work of LP is supported by the National Science and Engineering Research Council of Canada (NSERC). The work of AS is supported in part by the National Science Foundation under grant NSF-PHY0354990, by funds from Syracuse University and by Research Corporation.

%%%%%%%%%%%%%%%%%%%%%%%%%%%%%%%%%%%%%%%%%%%%%%%
\appendix
%%%%%%%%%%%%%%%%%%%%%%%%%%%%%%%%%%%%%%%%%%%%%%%

\section{Linear  Equations in Newtonian Gauge}
\label{Einstein_linear}

The evolution equations for a generic matter field are
\ba
\label{mattereom-generic1}
&&\delta'+\f{k}{aH}V-3(1+w)\f{k}{aH}\Phi'+3\l(\f{\delta P}{\delta \rho}-w\r)\delta=0\\
&&V'+(1-3w)V-\f{k}{aH}\l(\f{\delta P}{\delta\rho}-\f{\Pi}{\delta}\r)\delta -\f{k}{aH}(1+w)\Psi=0
\label{mattereom-generic2}
\ea
The linear Einstein equations are 
\begin{itemize}
\item {\it time-time}
\ba
\label{time-time} 
\nonumber
6\Psi+6\Phi'+2\f{k^2}{a^2H^2}\Phi =&& -\f{3E_i}{H^2}\delta_i-f_R\l(6\Psi+6\Phi'+2\f{k^2}{a^2H^2}\Phi\r)+3 f_{RR}\delta R'-f_R'\l(6\Psi+3\Phi'\r)+\nonumber\\
&&-\l[3f_{RR}\l(1+\f{H'}{H}\r)\delta R-\f{k^2}{a^2H^2}f_{RR}+3 f_{RR}'\r]\delta R
\ea
\item {\it momentum}
\ba
\label{momentum}
k\Psi+k\Phi'=\f{3a}{2H}E_i V_i-f_R\l(k\Psi+k\Phi'\r)+\f{1}{2}k(f_{RR}\delta R)'-\f{1}{2}k f_{RR}\delta R-\f{1}{2}kf_R'\Psi
\ea
\item {\it space-diagonal}
\ba
\label{trace}
\nonumber
&&\l(3+2\f{H'}{H}\r)\Psi-\f{1}{3}\f{k^2}{a^2H^2}\Psi+\Phi''+\l(1+\f{H'}{H}\r)\Phi'
+2\Phi'+\Psi'+\f{1}{3}\f{k^2}{a^2H^2}\Phi=\f{3E_i}{2H^2}\f{\delta P_i}{\delta\rho}\delta+\nonumber\\
&&-f_R\l[\l(3+2\f{H'}{H}\r)\Psi-\f{1}{3}\f{k^2}{a^2H^2}\Psi+\Phi''+\l(1+\f{H'}{H}\r)\Phi'+2\Phi'
+\Psi'+\f{1}{3}\f{k^2}{a^2H^2}\Phi\r]-\l(3+\f{H'}{H}\r)\f{f_{RR}\delta R}{2}+\nonumber\\
&&+\f{f_{RR}(\delta R)''}{2}+\l(1+\f{H'}{H}\r)\f{f_{RR}(\delta R)'}{2}+ \f{f_{RR}(\delta R)'}{2}+\f{1}{3}\f{k^2}{a^2H^2}f_{RR}\delta R-\Psi f_R''-\Psi\l(1+\f{H'}{H}\r)f_R'+\nonumber\\
&&-\f{f_R'}{2}(\Psi'+2\Psi+2\Phi')
\ea
\item {\it space off-diagonal}
\ba
\label{traceless}
\Phi-\Psi=\f{9}{2}\f{a^2}{k^2}E_i\Pi_i-f_R\l(\Phi-\Psi\r)+f_{RR}\delta R \ .
\ea
\end{itemize}
The linear perturbation to the Ricci scalar is
\be\label{deltaR}
\delta R=-2H^2\l[2\f{k^2}{a^2H^2}\Phi -\f{k^2}{a^2H^2}\Psi+3\Phi''+9\Phi'+3\Phi'+6\l(2+\f{H'}{H}\r)\Psi+3\l(1+\f{H'}{H}\r)\Phi'\r]
\ee

%%%%%%%%%%%%%%%%%%%%%%%%%%%%%%%%%%%%%%%%%%%%%%%%
\section{Mapping to the Einstein Frame}
\label{conformal}
%%%%%%%%%%%%%%%%%%%%%%%%%%%%%%%%%%%%%%%%%%%%%%%

Using the approaches of Chiba~\cite{Chiba:2003ir} and of Magnano \& Sokolowski~\cite{Magnano:1993bd}, following \cite{Cotsakis:1988}, we recast the gravitational action~(\ref{jordanaction}) into a dynamically equivalent form by introducing an intermediate scalar field $\Phi$.  The equivalent action is \cite{Magnano:1993bd}
\be
S=\frac{1}{2\kappa^{2}}\int d^4 x\sqrt{-g}\, \left[ (\Phi + f(\Phi)) + (1 + f_{\Phi})(R - \Phi) \right] +\int d^4 x\sqrt{-g}\, {\cal L}_{\rm m}[\chi_i,g_{\mu\nu}] \ ,
\label{IntermediateAction}
\ee
where $f_{\Phi} \equiv \partial{f}/\partial{\Phi}$.  One can verify that, if $d^{2}f/d\Phi^{2} \neq 0$, the field equation for $\Phi$ is $R=\Phi$, which reduces~(\ref{IntermediateAction}) to the original action. Next consider the conformal transformation
\be
{\tilde g}_{\mu\nu} = e^{2 \omega(x^{\alpha})} g_{\mu\nu} \ ,
\label{conftrans}
\ee
such that the function $\omega(x^{\alpha})$ satisfies
\be
e^{-2 \omega}(1+f_{R}) = 1 \ .
\label{constraint}
\ee
With this choice of $\omega$ the action~(\ref{IntermediateAction}) transforms into an action with the usual Hilbert-Einstein form for gravity. If we now define the scalar field  $\phi \equiv 2\omega / \beta\kappa$, where $\beta \equiv\sqrt{2/3}$, the resulting action becomes 
\be
{\tilde S}=\frac{1}{2\kappa^{2}}\int d^4 x\sqrt{-{\tilde g}}\, {\tilde R} 
+\int d^4 x\sqrt{-{\tilde g}}\, 
\left[-\frac{1}{2}{\tilde g}^{\mu\nu}(\tilde{\nabla}_{\mu}\phi)\tilde{\nabla}_{\nu}\phi -V(\phi)\right] +\int d^4 x\sqrt{-{\tilde g}}\, e^{-2\beta\kappa\phi} {\cal L}_{\rm m}[\chi_i,e^{-\beta\kappa\phi}{\tilde g}_{\mu\nu}]\ ,
\label{einsteinaction}
\ee
where the potential $V(\phi)$ is determined entirely by the original form~(\ref{jordanaction}) of the action and is given by
\be
V(\phi)=\frac{1}{2\kappa^{2}}\frac{R f_{R} - f}{(1+f_{R})^{2}} \ .
\label{einsteinpotential}
\ee
The Einstein-frame line element can be written in familiar FRW form as
\be
d{\tilde s}^2 ={\tilde a}^2(-d{\tau}^2+d{\bf x}^2) \ ,
\label{einsteinFRWmetric} 
\ee
where the Jordan and Einstein metrics are related through the conformal transformation ${\tilde a}^2\equiv e^{\beta\kappa\phi} \,a^2$. It is 
also convenient to define an Einstein-frame matter energy-momentum tensor by
\be
{\tilde T}_{\mu\nu} = ({\tilde \rho}+ {\tilde P}){\tilde U}_{\mu} {\tilde U}_{\nu} + {\tilde P} {\tilde g}_{\mu\nu}\ ,
\label{einsteinperfectfluid}
\ee
where ${\tilde U}_{\mu}\equiv e^{\beta\kappa\phi/2} \,U_{\mu}$, ${\tilde \rho}\equiv e^{-2 \beta\kappa\phi} \rho$ and ${\tilde P}\equiv e^{-2 \beta\kappa\phi} P$.

The equations of motion obtained by varying the action with respect to the metric ${\tilde g}_{\mu\nu}$ are
\be
{\tilde G}_{\mu\nu} = 8 \pi G \tilde{T}_{\mu \nu} + \frac{1}{2} \tilde{\nabla}_{\mu} \phi \tilde{\nabla}_{\nu} \phi + \frac{1}{2}(\tilde{g}^{\alpha \gamma} \tilde{\nabla}_{\alpha} \phi \tilde{\nabla}_{\gamma} \phi ) \tilde{g}_{\mu\nu} - V(\phi) \tilde{g}_{\mu\nu} \ ,
\label{einsteineom}
\ee
and are more familiar than those in the Jordan frame, although there are some crucial distinctions.  Most notably, in this frame test matter particles do not freely fall along geodesics of the metric ${\tilde g}_{\mu\nu}$, since the scalar field is also coupled to matter.  

The remaining equations of motion, for the scalar field and for the perfect fluid matter, are given respectively by
\bea
\ddot{\phi} + 2\tilde{\hub}\dot{\phi} + \tilde a^{2}V_{\phi} = \f{1}{2}\kappa\beta \tilde a^{2}(\tilde{\rho} - 3 \tilde{P} ) \ ,
\label{scalareom}
\\
\dot{\tilde{\rho}} +3\tilde{\hub}(\tilde{\rho} + \tilde{P}) = - \f{1}{2}\kappa\beta \dot{\phi}( \tilde\rho-3\tilde P) \ .
\eea
where  $V_{\phi}= dV/d\phi$.

At linear level in the perturbations, the evolution equations for scalar field and CDM perturbations are
\ba\label{phisecond}
&&\delta\phi''+\l(3+\f{\tilde{H}'}{\tilde{H}}\r)\delta\phi'+\f{k^2}{\tilde{a}^2\tilde{H}^2}\delta\phi
-4(\tilde{\Psi}'+\tilde{\Psi})\tilde{\Psi}'+\f{m_{\phi}^2}{\tilde{H}^2}\delta\phi
=\f{3}{2}\f{\beta}{\kappa}\f{\tilde{\Omega}}{\tilde{H}^2}\tilde{\delta}_m\\
\label{cdmsecond-Einstein}
&&\tilde{\delta}_m''+\l(2+\f{\tilde{H}'}{\tilde{H}}-\f{\beta\kappa\phi'}{2}\r)\tilde{\delta}_m'
-3\tilde{\Phi}''-3\l(2+\f{\tilde{H}'}{\tilde{H}}-\f{\beta\kappa\phi'}{2}\r)\tilde{\Phi}'
+\f{\beta\kappa}{2}\l[\delta\phi''+\l(2+\f{\tilde{H}'}{\tilde{H}}
-\f{\beta\kappa\phi'}{2}\r)\delta\phi'\r]+\nonumber\\
&&+\f{k^2}{\tilde{a}^2\tilde{H}^2}\l(\tilde{\Psi}
-\f{\beta\kappa\delta\phi}{2}\r)=0\nonumber\\
\ea
The mapping prescription at the linear level is the following
\ba\label{linear_mapping}
&&\tilde{\Psi}=\Psi+\f{\beta\kappa\delta\phi}{2},\hspace{1cm}\tilde{\Phi}=\Phi-\f{\beta\kappa\delta\phi}{2}\\
&&\tilde{\delta}=\delta-2\beta\kappa\delta\phi,\hspace{1cm}\tilde{\delta P}=\delta P - 2\beta\kappa\delta\phi\\
&&\tilde{v}=v,\hspace{2.5cm}\tilde{\Pi}=\Pi
\ea

%%%%%%%%%%%%%%%%%%%%%%%%%%%%%%%%%%%%%%%%%%%%%%%

\end{document}